\documentclass[prd,nofootinbib,notitlepage]{revtex4-1}
\usepackage{epsfig,amsmath,amssymb,slashed}
\usepackage{graphicx}
\usepackage{hyperref}
\usepackage{color}

\newcommand{\di}{{\rm d}}
\newcommand{\ii}{i}

\def\wT{{\widehat T}}
\def\wj{{\widehat j}}

\def\wrho{{\widehat{\rho}}}
\def\wrhol{{\widehat\rho_{\rm LE}}}

\def\wA{\widehat A}
\def\wB{\widehat B}

\newcommand{\tr}{{\rm tr}}  
  
\newcommand{\e}{{\rm e}}

\newcommand{\x}{{\rm x}}

\newcommand{\subb}{{\beta(x)}}
\newcommand{\be}{\begin{equation}}
\newcommand{\ee}{\end{equation}}                                                                               
\newcommand{\bea}{\begin{eqnarray}}
\newcommand{\eea}{\end{eqnarray}}                                                                               
\begin{document}

\title{Reworking the Zubarev's approach to non-equilibrium quantum statistical 
mechanics} 

\author{F. Becattini}
\affiliation{Universit\`a di Firenze and INFN Sezione di Firenze, Florence, Italy}
\author{M. Buzzegoli}
\affiliation{Universit\`a di Firenze and INFN Sezione di Firenze, Florence, Italy}
\author{E. Grossi}
\affiliation{Institut für Theoretische Physik, University of Heidelberg, Heidelberg,
Germany}

\begin{abstract}
In this work the non-equilibrium density operator approach introduced by Zubarev 
more than 50 years ago to describe quantum systems at local thermodynamic
equilibrium is revisited. This method - which was used to obtain the first ''Kubo" 
formula of shear viscosity, is especially suitable to describe quantum effects in
fluids. This feature makes it a viable tool to describe the physics of the Quark 
Gluon Plasma in relativistic nuclear collisions. 
\end{abstract}

\maketitle

\section{Introduction}
\label{intro}

One of the authors (F.B.) would like to start this paper with a personal recollection. 
I (F.B.) first ran across Zubarev's papers when I was studying the derivation by A. Hosoya 
{\it et al} \cite{hosoya} of the shear viscosity in quantum field theory, a result 
widespreadly known as ``Kubo formula", like many of the same sort. This derivation 
was overtly based on the Zubarev's method of non-equilibrium density (or statistical) 
operator and I surmised that this method must have been a very important and renowned 
tool in quantum statistical mechanics. 
In fact, surprisingly, it could be hardly found in textbooks as well as in recent
literature and I did not quite understand why the founding method of such an important 
formula was that overlooked. After some more self-education I realized that, perhaps, 
part of the problem was that Zubarev himself did not put the right emphasis on the 
crucial feature that his proposed operator should possess: to be {\em stationary}, 
hence well suited to be used in relativistic quantum field theory as a density 
operator in the Heisenberg representation. 
A non-equilibrium stationary density operator sounds somewhat contradictory, but 
indeed this is not the case if we deal with a system which, at some time, is known 
to be in local thermodynamic equilibrium, as we will see in more detail in 
Section~\ref{densop}.

In this work, we would like not just to summarize Zubarev's method \cite{zuba1,zuba2,zuba3}, 
rather to make a critical appraisal and to provide a reformulation thereof which 
highlights the nice features of this approach in a hopefully clear fashion. I also 
hope that this work will contribute to do justice to Zubarev and his remakable achievement.

\subsection*{Notation}

In this paper we use the natural units, with $\hbar=c=K=1$.\\ 
The Minkowskian metric tensor is ${\rm diag}(1,-1,-1,-1)$; for the Levi-Civita
symbol we use the convention $\epsilon^{0123}=1$.\\  
Operators in Hilbert space will be denoted by a large upper hat, e.g. $\wT$ while unit 
vectors with a small upper hat, e.g. $\hat v$. Scalar products and contractions
are sometimes denoted with a dot, e.g. $A_\mu B^\mu = A \cdot B$.

\section{Local thermodynamic equilibrium}
\label{lte}

The Zubarev formalism can be used in non-relativistic as well as in relativistic 
quantum statistical mechanics. We can then start at once from the latter, more
general case, which is applicable to relativistic fluids out of equilibrium \cite{rischke}. 
The relativistic version of the non-equilibrium density operator was first put forward
by Zubarev himself and his collaborators in 1979 \cite{zubarel} and later reworked
by Van Weert in ref.~\cite{weert}. 

The starting point is the definition of the {\em local equilibrium} density operator.
In relativity, this notion needs the specification of a one-parameter family of 
3D space-like hypersurfaces $\Sigma(\tau)$ (see fig.~\ref{foliation}), also known
as foliation of the spacetime \cite{zubarel,weert,betaframe,hongo}. The "time" 
$\tau$ does not necessarily coincide with the proper time marked by comoving clocks. 
The local equilibrium density operator $\wrhol$ is obtained by maximizing the total 
entropy:
\be
  S = -\tr(\wrho \log (\wrho))
\ee
with constrained values of energy-momentum and charge density, which should be equal
to the actual values. In a covariant formulation, these densities are obtained by 
projecting the mean values of the stress-energy tensor and current onto the normalized 
vector perpendicular to $\Sigma$:
\be\label{constr}
 n_\mu \tr \left(\wrho \, \wT^{\mu\nu}\right) = n_\mu T^{\mu\nu}, 
 \qquad n_\mu \tr \left(\wrho \, \wj^{\mu}\right) = n_\mu j^{\mu}.
\ee
where $T^{\mu\nu}$ and $j^\mu$ are the true values of the stress-energy and current
fields. The operators in eq.~(\ref{constr}) are in the Heisenberg representation. 
In addition to the energy, momentum, and charge densities, one should include the angular 
momentum density, but if the stress-energy tensor is the Belinfante \cite{becaflor}
this further constraint is redundant and can be disregarded. 

The resulting operator is the Local Equilibrium Density Operator (LEDO):
\be\label{ledo} 
  \wrhol = \dfrac{1}{Z_{\rm LE}} 
  \exp \left[ -\int_{\Sigma(\tau)} \di \Sigma \; n_\mu \left( \wT^{\mu\nu}(x) 
  \beta_\nu(x) - \zeta(x) \wj^\mu(x) \right) \right]
\ee  
where $\beta$ and $\zeta$ are the relevant Lagrange multiplier functions for this 
problem, whose meaning is the four-temperature vector and the ratio between local 
chemical potential and temperature, respectively \cite{betaframe}. The $\di \Sigma$
is the measure of the hypersurface induced by the Minkowskian metric, and the fields 
$\beta$ and $\zeta$ are the solution of the contraints (\ref{constr}) with $\wrho 
= \wrhol$, namely:
\be\label{solutions}
 n_\mu \tr \left(\wrhol \, \wT^{\mu\nu}\right) = n_\mu T_{\rm LE}^{\mu\nu}
 [\beta,\zeta,n] = n_\mu T^{\mu\nu}, 
 \qquad n_\mu \tr \left(\wrhol \, \wj^{\mu}\right) = n_\mu j_{\rm LE}^{\mu}
 [\beta,\zeta,n] = n_\mu j^{\mu}. 
\ee
These equations indeed define a vector field $\beta$ which in turn can be used
as a hydrodynamic frame, the $\beta$ \cite{betaframe} or thermodynamic frame
\cite{kovtun}, by identifying the four-velocity with:
\be\label{ubeta}
  u = \frac{\beta}{\sqrt{\beta^2}} = T u
\ee
which somehow inverts the usual definition.

It is important to stress that the LEDO in the eq.~(\ref{ledo}) is {\em not} stationary 
because the operators are generally time dependent. The sufficient condition for 
the stationarity is that $\beta$ is a Killing vector field and $\zeta$ a constant 
and, in this case, the LEDO becomes the general global thermodynamic equilibrium 
operator \cite{becacov}.

\section{Non-equilibrium density operator revisited}
\label{densop}

\begin{center}
\begin{figure}[ht]
\includegraphics[width=0.45\textwidth]{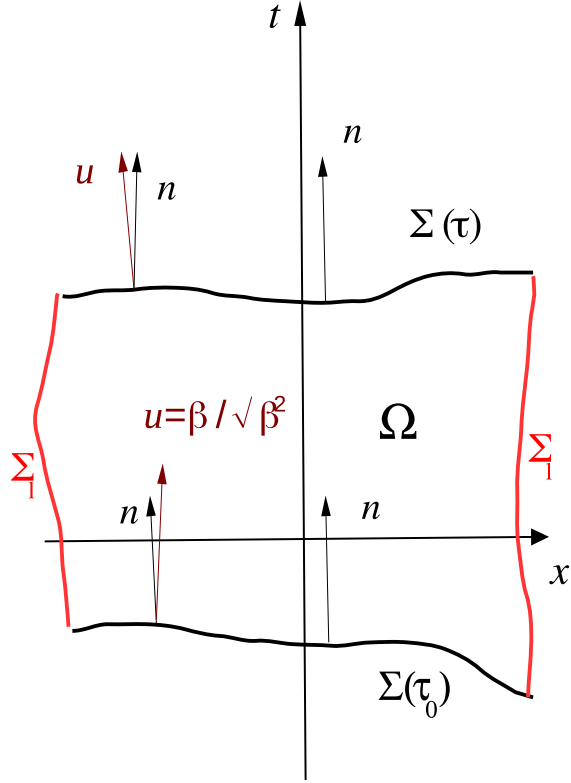}
\caption{Spacelike hypersurfaces $\Sigma(\tau)$, $\Sigma(\tau_0)$ and their normal 
unit vector $n$ defining local thermodynamical equilibrium for a relativistic fluid 
in Minkwoski spacetime. At the timelike boundary $\Sigma_l$ the flux is supposed
to vanish.} 
\label{foliation}
\end{figure}
\end{center}

The true density operator in the Heisenberg representation must be stationary by
definition, whereas the LEDO is not. How to work it out?  
The solution (which is an amendment of Zubarev's original idea) is overly simple: 
if, at some initial time $\tau_0$ the system is known to be in local thermodynamic 
equilibrium, the actual, stationary, non-equilibrium density operator (NEDO) is 
$\wrhol(\tau_0)$. Therefore, the true mean values of quantum operators should be 
calculated as:
$$
 \langle \widehat O \rangle \equiv \tr (\wrho \widehat O) = \tr (\wrhol(\tau_0) 
\widehat O)
$$
One can rewrite $\wrhol(\tau_0)$ in terms of the operators at the present "time" 
$\tau$ by means of the Gauss' theorem, taking into account that $\wT$ and $\wj$ 
are conserved. Defining:
$$
  \di \Sigma_\mu = \di \Sigma \, n_\mu
$$
and $\di \Omega$ being the measure of a 4D region in spacetime, we have
\be\label{gauss}
- \int_{\Sigma(\tau_0)} \!\!\!\!\!\! \di \Sigma_\mu \; \left( \wT^{\mu\nu} 
  \beta_\nu - \wj^\mu \zeta \right) = 
 - \int_{\Sigma(\tau')} \!\!\!\!\!\! \di \Sigma_\mu \; \left( \wT^{\mu\nu} 
  \beta_\nu - \wj^\mu \zeta \right) 
  + \int_\Omega \di \Omega \; \left( \wT^{\mu\nu} \nabla_\mu \beta_\nu - \wj^\mu 
 \nabla_\mu \zeta \right), 
\ee
where $\nabla$ is the covariant derivative. The region $\Omega$ is the portion of spacetime 
enclosed by the two hypersurface $\Sigma(\tau_0)$ and $\Sigma(\tau)$ and the timelike 
hypersurface at their boundaries, where the flux of ($\wT^{\mu\nu} \beta_\nu(x) - 
\wj^\mu \zeta(x)$) is supposed to vanish (see fig.~\ref{foliation}). Consequently, 
the stationary NEDO reads:
\be\label{nedo}
 \wrho =  \dfrac{1}{Z} 
 \exp\left[ - \int_{\Sigma(\tau_0)} \!\!\!\!\!\! \di \Sigma_\mu \; \left( \wT^{\mu\nu} 
  \beta_\nu - \wj^\mu \zeta \right) \right] =
 \dfrac{1}{Z} 
 \exp\left[ - \int_{\Sigma(\tau)} \!\!\!\!\!\! \di \Sigma_\mu \; \left( \wT^{\mu\nu} 
  \beta_\nu - \wj^\mu \zeta \right) + \int_\Omega \di \Omega \; \left( \wT^{\mu\nu} 
  \nabla_\mu \beta_\nu - \wj^\mu \nabla_\mu \zeta \right) \right] 
\ee
This expression is the generally covariant form of the one used in ref.~\cite{hosoya}
(Equation (2.9) therein) with the only difference that the factor $\exp[\varepsilon(t-\tau)]$
does not appear in the second term. We will see in Section \ref{kubo} that such
a factor is not necessary to obtain the correct ``Kubo" formulae.

The NEDO can be worked out perturbatively by identifying the two terms in the 
exponent of (\ref{nedo}): 
\be\label{aa}
 \wA = - \int_{\Sigma(\tau)} \!\!\!\!\!\! \di \Sigma_\mu \; 
  \left( \wT^{\mu\nu} \beta_\nu - \wj^\mu \zeta \right)
\ee
and:
\be\label{bb}
  \wB = \int_{\Omega} \di \Omega \; \left( \wT^{\mu\nu} 
  \nabla_\mu \beta_\nu - \wj^\mu \nabla_\mu \zeta \right) 
\ee  
and assuming that $\widehat B$ is small compared to $\wA$; this happens if the system
has small correlation length and if the gradients in (\ref{bb}) are small, that
is the hydrodynamic limit. We can then use the identity:
$$
 \exp[\wA+\wB] = \exp[\wA] + \int_0^1 \di z \; \exp[z(\wA+\wB)] \wB \exp[-z\wA]
 \exp[\wA]
$$
The expansion of $\exp[\wA+\wB]$ can be iterated in the integrand and one obtains
an operator expansion in $\wB$. Taking into account that 
$$
 Z = \tr(\exp[\wA+\wB])
$$
at the lowest order in $\wB$ (linear response):
\be\label{lrt}
 \wrho \simeq \wrhol + \int_0^1 \di z \; \exp[z\wA] \wB \exp[-z\wA] \wrhol 
 - \langle \wB \rangle_{\rm LE} \wrhol
\ee
which is the starting point to obtain the ``Kubo" formulae.

It should be pointed out that the original Zubarev formulae were somewhat different 
\cite{zubarel}. We will work it by using Cartesian coordinates and hyperplanes 
as hypersurfaces. Zubarev modified the equation for the NEDO in the Heisenberg 
representation:
\be\label{zubeq}
  \frac{\di \wrho}{\di t} = -\varepsilon(\wrho - \wrhol)
\ee
being $\varepsilon > 0$ a real parameter whose limit $\varepsilon \to 0$ is to be 
taken {\em after} the thermodynamic limit. The solution of the above equation at
the present time, which can be chosen to be $t=0$, reads:
\be\label{solution}
  \wrho(0) = \wrhol - \int^0_{-\infty} \di t \; \e^{\varepsilon t} 
 \frac{\di \wrhol}{\di t} 
\ee
One can now use the general expression for the derivative of an exponential to
calculate:
$$
  \frac{\di \e^{\wA}}{\di t} = \int_0^1 \di z \; \e^{z\wA} \frac{\di \wA}{\di t} 
  \e^{(1-z)\wA} 
$$
with $\wA$ given by the equation (\ref{aa}). This implies:
$$
  \frac{\di Z_{\rm LE}}{\di t} = \frac{\di}{\di t} \tr (\e^{\wA}) = 
  \tr \left( \frac{\di \wA}{\di t} \e^{\wA} 
  \right) = Z_{\rm LE} \langle \frac{\di \wA}{\di t} \rangle_{\rm LE} 
$$
so that:
\be\label{drholdt}
 \frac{\di \wrhol}{\di t} = \int_0^1 \di z \; \e^{z\wA} \frac{\di \wA}{\di t} 
  \e^{-z\wA} \wrhol - \langle \frac{\di \wA}{\di t} \rangle_{\rm LE} \wrhol  
\ee
If the surface boundary terms vanish, we have
\be\label{dadt}
  \frac{\di \wA}{\di t} = - \int \di^3 {\x} \; \frac{\partial}{\partial t}
  (\wT^{0\nu} \beta_\nu) = - \int \di^3 {\x} \; \partial_\mu (\wT^{\mu\nu} 
  \beta_\nu) = - \int \di^3 {\x} \; \wT^{\mu\nu} \partial_\mu \beta_\nu
\ee
By plugging the (\ref{dadt}) and the (\ref{drholdt}) into the (\ref{solution}),
we have:
$$
 \wrho(0) - \wrhol = \int_0^1 \di z \; \e^{z\wA} \int^0_{-\infty} \di^4 x \; 
\; \e^{\varepsilon t} \wT^{\mu\nu} \partial_\mu \beta_\nu \e^{-z\wA} \; \wrhol
  - \int^0_{-\infty} \di^4 x \; \; \e^{\varepsilon t} \langle \wT^{\mu\nu} \rangle_{\rm LE}
  \partial_\mu \beta_\nu \; \wrhol 
$$
Taking into account (\ref{bb}), the above equation is basically the linear approximation 
(\ref{lrt}) with an extra factor $\exp(\varepsilon t)$ in the integrand. 
In a sense, the Zubarev assumption (\ref{zubeq}) of a small source term in the 
density operator evolution equation in the Heisenberg representation leads to the 
linear approximation of the fully stationary density operator operator (\ref{nedo}).
However, it should be emphasized that such an extra factor is not necessary. The
Heisenberg equation for the true density operator is $\di \wrho/\di t =0$ does 
not need any modification for the derivation of the Kubo formulae or any other
result depending on local thermodynamic equilibrium, as we will show in Sect.~\ref{kubo}. 
A fully relativistic viewpoint with the application of the Gauss theorem makes 
the derivation of the NEDO expression (\ref{nedo}) straightforward, transparent
and simple. 

\section{Entropy production}
\label{entropy}

A remarkable consequence of this approach is the derivation of a general equation
for the entropy production rate, which was reported in refs.~\cite{zubarel,weert}. 
Let us start with the assumption that $S$ is an integral of an entropy current 
$s^\mu$:
$$
   S(\tau) = - \tr (\wrhol(\tau) \log \wrhol(\tau)) = \int_{\Sigma(\tau)} 
  \di \Sigma_\mu \; s^\mu
$$
On the other hand, the entropy can be expanded by using the (\ref{ledo}):
\begin{align}\label{entr2}
   S(\tau) = \tr (\wrhol(\tau) \log \wrhol(\tau)) &= \log Z_{\rm LE} + 
   \int_{\Sigma(\tau)} \di \Sigma_\mu \; \langle \wT^{\mu\nu}\rangle_{\rm LE} 
   \beta_\nu - \zeta \langle \wj^{\mu} \rangle_{\rm LE} \nonumber \\
   &= \log Z_{\rm LE} + 
   \int_{\Sigma(\tau)} \di \Sigma_\mu \; \left( T^{\mu\nu} \beta_\nu - 
    \zeta j^\mu \right)
\end{align}
where we have used the constraints (\ref{constr}), taking into account that 
$\di \Sigma_\mu = \di \Sigma \; n_\mu$.

The derivative with respect to $\tau$ can be computed by taking advantage of a
general expression for the variation of an integral between two infinitesimally
closed hypersurfaces:
\be\label{entrate}
  \frac{\di S}{\di \tau} = \int_{\Sigma(\tau)} \di \Sigma (n \cdot U) 
   \nabla \cdot s + \frac{1}{2} \int_{\partial\Sigma(\tau)} \di \tilde S_{\mu\nu} 
   (s^\mu U^\nu - s^\nu U^\mu)
\ee
where $\partial \Sigma$ is the 2D boundary of $\Sigma$ and  $U^\mu=\partial x^\mu/\partial \tau$;
the $\tilde S$ is the dual of the surface element. Now, assume that the boundary 
term does not contribute and calculate the same derivative by using the expression
(\ref{entr2}):
\begin{align}\label{entrate2}
  \frac{\di S}{\di \tau} &= \frac{\di \log Z_{\rm LE}}{\di \tau} +
  \int_{\Sigma(\tau)} \di \Sigma (n \cdot U) \nabla_\mu \left( T^{\mu\nu} \beta_\nu - 
    \zeta j^\mu \right) \nonumber \\
  &= \frac{\di \log Z_{\rm LE}}{\di \tau} +
  \int_{\Sigma(\tau)} \di \Sigma (n \cdot U) T^{\mu\nu} \nabla_\mu \beta_\nu - 
    j^\mu \nabla_\mu \zeta
\end{align}
where we have taken advantage of the conservation of the {\em exact} values $T^{\mu\nu}$
and $j^\mu$. The remaining task is to calculate the derivative of $\log Z_{\rm LE}$,
which can be done by using its definition:
$$
  \frac{\di \log Z_{\rm LE}}{\di \tau} = \frac{1}{Z_{\rm LE}}
  \frac{\di}{\di \tau} \tr (\exp[\wA])
$$
with $\wA$ in eq.~(\ref{aa}). By using the same formula of the derivative of
a $\tau$-dependent integral in eq.~(\ref{entrate2}) and assuming that the boundary
term vanishes:
\be\label{dlogz}
  \frac{1}{Z_{\rm LE}} \frac{\di}{\di \tau} \tr (\exp[\wA]) =  
  \frac{1}{Z_{\rm LE}} \tr \left( \frac{\di \wA}{\di \tau} 
  \exp[\wA] \right) = \langle \frac{\di \wA}{\di \tau} \rangle_{\rm LE} = 
  - \int_{\Sigma(\tau)} \di \Sigma (n \cdot U) \left( T^{\mu\nu}_{\rm LE} 
   \nabla_\mu \beta_\nu - j^\mu_{\rm LE} \nabla_\mu \zeta \right)
\ee
By plugging (\ref{dlogz}) into the (\ref{entrate2}) and comparing with (\ref{entrate}),
taking into account that the equation should hold for any $\tau$ we have:
\be\label{entratef}
  \nabla \cdot s = ( T^{\mu\nu} - T^{\mu\nu}_{\rm LE}) \nabla_\mu \beta_\nu
  - (j^\mu - j^\mu_{\rm LE}) \nabla_\mu \zeta
\ee
which was found in ref.~\cite{zubarel} and tells us that the deviations of the 
conserved currents actual values from those at local thermodynamic equilibrium 
are responsible for the entropy production.

\section{Kubo formulae}
\label{kubo}

Let us now apply the expansion of the NEDO (\ref{lrt}) to calculate the mean value 
of a local operator $\widehat O$ at the present time $t$:
\be\label{kuboexp}
  \langle \widehat O (x) \rangle \simeq \langle \widehat O (x) \rangle_{\rm LE} 
  - \langle \widehat O (x) \rangle_{\rm LE} \langle \widehat B \rangle_{\rm LE}
  + \int_0^1 \di z \; \langle \widehat O (x) \e^{z\widehat A} \widehat B 
  \e^{-z\widehat A} \rangle_{\rm LE}
\ee
where $\widehat A$ and $\widehat B$ are in eq.~(\ref{aa}) and (\ref{bb}) respectively.
To work out the formula (\ref{kuboexp}) it is customary to approximate the 
$\widehat A$ in the $z$ integral on the right hand side with the global equilibrium 
expression. In a covariant fashion, this means making a zero-order approximation 
of the Taylor expansion of the thermodynamic fields from the point $x$ where the 
operator $\widehat O$ is to be calculated:
\be\label{ltetoeq}
\widehat A = - \int_{\Sigma(\tau)} \!\!\!\!\!\! \di \Sigma_\mu \; 
 \left( \wT^{\mu\nu} \beta_\nu - \wj^\mu \zeta \right) \simeq 
 - \beta_\nu(\tau,\sigma) \int_{\Sigma(\tau)} \!\!\!\!\!\! \di \Sigma_\mu \; 
 \wT^{\mu\nu} + \zeta(\tau,\sigma) \int_{\Sigma(t)} \!\!\!\!\!\! \di \Sigma_\mu 
 \; \wj^\mu = - \beta_\nu(x) \widehat P^\nu + \zeta(x) \widehat Q
\ee
where $\widehat P$ is the total four-momentum and $\widehat Q$ the total charge.
Hence:
\be\label{eqlbrm}
  \wrhol \simeq \dfrac{1}{Z_{\rm LE}} \exp[-\widehat A] \simeq \dfrac{1}{Z} 
  \exp[-\beta(x) \cdot \widehat P + \zeta(x) \widehat Q] \equiv \wrho_{\rm eq(x)}
\ee  
that is, $\wrho_{\rm eq(x)}$ is the global equilibrium density operator having as
constant inverse temperature four-vector the same vector at the point $x$ and
similarly for $\zeta$. 

Furthermore, we will replace the integration region enclosed by the two LTE 
hypersurfaces at $t$ and $t_0$ with the spacelike tangent hyperplanes at 
the points $x=(\tau,\sigma)$ and $x_0 = (\tau_0,\sigma)$ respectively, whose normal 
versor is $n$. This allows to carry out the integration over Minkowski spacetime
by using Cartesian coordinates, that is the time $t$ marked by an observer moving 
with velocity $n$, and a vector of coordinates ${\bf x}$ for the hyperplanes.
These approximations make it possible to replace covariant derivatives with usual 
partial derivatives in Cartesian coordinates:
\be\label{cartesianize}
  \int_\Omega \di \Omega \; \left( \wT^{\mu\nu} \nabla_\mu \beta_\nu - \wj^\mu 
  \nabla_\mu \zeta \right) \rightarrow \int_{T\Omega} \di^4 x \; \left( \wT^{\mu\nu} 
  \partial_\mu \beta_\nu - \wj^\mu \partial_\mu \zeta \right) 
\ee
where $T\Omega$ is the region encompassed by the two hyperplanes. Thereby, and 
provided that $n(x) = \hat \beta(x)$, that is that the local equilibrium hypersurface
is locally normal to the flow velocity defined by the four-temperature vector 
\cite{betaframe}, the formula (\ref{kuboexp}) can be turned into a more 
manageable one (see Appendix A for a summary of the derivation) involving the 
commutators of the operator $\widehat O$ with the stress-energy tensor and the current 
operators:
\be\label{ltexpa}
  \langle \widehat O (x) \rangle - \langle \widehat O (x) \rangle_{\rm LE} 
  \simeq \ii T \int_{t_0}^{t} \!\!\! \di^4 x^\prime   
  \int_{t_0}^{t^\prime} \!\!\! \di \theta \; \left(\langle 
  [\widehat O(x),\wT^{\mu\nu}(\theta,{\bf x}^\prime)] \rangle_\subb \partial_\mu 
  \beta_\nu(x^\prime) - \langle [\widehat O(x),\wj^\mu(\theta,{\bf x}^\prime)] 
  \rangle_\subb \partial_\mu \zeta(x^\prime) \right)
\ee
where $T = 1/\sqrt{\beta^2}$ and the subscript $\subb$ stands for averaging with 
the density operator in eq.~(\ref{eqlbrm}). It is important to stress the different 
time arguments for the operators and the thermodynamic fields in (\ref{ltexpa}).

From eq.~(\ref{ltexpa}) it turns out that the deviation from LTE of the mean value
of $\widehat O$ at any time depends on the whole history of the thermodynamic fields
$\beta$ and $\zeta$. However, the correlation length between $\widehat O (x)$ and 
both $\wT(x'),\wj(x')$ is typically much smaller than the distance over which the 
gradient of $\beta$ and $\zeta$ have significant variations. This statement amounts 
to assume a separation between the typical microscopic interaction scale and the 
macroscopic hydrodynamical scale. One would then be tempted to take the gradients 
out of the integral in eq.~(\ref{ltexpa}). However, much care should be taken in 
this because one should keep in mind that the derivation of the formula (\ref{ltexpa}), 
more precisely the non-equilibrium density operator (\ref{nedo}), required the 
vanishing of the flux of $\wT^{\mu\nu} \beta_\nu - \wj^\mu \zeta $ at the boundary 
timelike hypersurface. If one expands the perturbation of the thermodynamic fields 
with respect to their equilibrium value, by definition those at the point $x$, 
that is:
$$
 \delta\beta \equiv \beta - \beta_{\rm eq} = \beta - \beta(x) 
 \qquad \delta\zeta \equiv \zeta - \zeta_{\rm eq} = \zeta - \zeta(x)
$$ 
in Fourier series, the only relevant components in the hydrodynamical limit for the
integral (\ref{ltexpa}) are those with very small frequency $\omega$ and wave-vector 
${\bf k}$. At the same time, the vanishing of the flux can be achieved by enforcing
periodicity of the perturbations in ${\bf x}-{\bf x'}$. Taking these requirements 
into account, the perturbations will include only the {\em smallest} wave four-vector 
$K$:
\be\label{pertu}
   \delta \beta_\nu(x') \simeq A_\nu \dfrac{1}{2\ii} (\e^{\ii K \cdot (x'-x)} - 
   \e^{-\ii K \cdot (x'-x)})
\ee
being $A_\nu$ a real constant, the amplitude of the smallest wave four-vector 
Fourier component. The above form fulfills $\delta\beta(x')=0$ as well as the request 
of vanishing flux provided that $K^i = \pi/L_i$ being $L_i$ the size of the compact 
domain in the direction $i$. Hence, after the us of (\ref{pertu}), the limit $K \to 0$
is to be taken, which is equivalent to the limit of infinite volume. 
The gradient of the (\ref{pertu}) (keep in mind that in eq.~(\ref{ltexpa}) 
$\partial_\mu = \partial/\partial x'^\mu$) can then be written as: 
\be\label{pertu2}
  \partial_\mu \beta_\nu \simeq K_\mu A_\nu \dfrac{1}{2} (\e^{\ii K \cdot (x'-x)} + 
   \e^{-\ii K \cdot (x'-x)}) = \partial_\mu \beta_\nu (x) {\rm Re} \; 
   \e^{-\ii K \cdot (x'-x)} = {\rm Re} \; \partial_\mu \beta_\nu (x)
   \e^{-\ii K \cdot (x'-x)} 
\ee
Plugging the (\ref{pertu2}) in the (\ref{ltexpa}), in the limit $K \to 0$, 
one obtains:
\bea\label{ltexpa2}
  \langle \widehat O (x) \rangle - \langle \widehat O (x) \rangle_{\rm LE} 
  \simeq && \partial_\mu \beta_\nu (x) \lim_{K \to 0} {\rm Im} \; T 
  \int_{t_0}^{t} \!\!\! \di^4 x' \int_{t_0}^{t^\prime} \!\!\! \di \theta \; 
  \langle [\wT^{\mu\nu}(\theta,{\bf x}'),\widehat O(x)] \rangle_\subb
  \e^{-\ii K \cdot (x'-x)} \nonumber \\
  && - \partial_\mu \zeta (x) \lim_{K \to 0} {\rm Im} \; T \int_{t_0}^{t} 
   \!\!\! \di^4 x' \int_{t_0}^{t'} \!\!\! \di \theta \langle 
  [\wj^\mu(\theta,{\bf x}'),\widehat O(x)] \rangle_\subb 
  \e^{-\ii K \cdot (x'-x)}  
\eea
As the macroscopic time scale $t-t_0$ and the microscopic time scale inherent in
the correlators are so different, one can take the limit $t_0 \to -\infty$.
If the functions:
$$
  \int \di^3 \x' \; \langle [\widehat X (\theta,{\bf x}'),\widehat O(x)] \rangle_\subb
$$
with $\widehat X = \wT, \wj$ remain finite for $\theta \to -\infty$, then the 
eq.~(\ref{ltexpa2}), after integration by parts in $t^\prime$ can be turned into:
\bea\label{ltexpa3}
  \langle \widehat O (x) \rangle - \langle \widehat O (x) \rangle_{\rm LE} 
  \simeq && \; \partial_\mu \beta_\nu (x) n^\alpha \dfrac{\partial}{\partial K^\alpha} 
  \Big|_{n \cdot K=0}
  \lim_{K_T \to 0} {\rm Im} \; \ii T \int_{-\infty}^{t} \!\!\! \di^4 x^\prime \;
  \langle [\widehat O(x),\wT^{\mu\nu}(x')] \rangle_\subb \e^{-\ii K \cdot (x'-x)} 
  \nonumber \\
  && - \partial_\mu \zeta (x) n^\alpha \dfrac{\partial}{\partial K^\alpha} 
  \Big|_{n \cdot K=0} {\rm Im} \; \ii T \int_{-\infty}^{t} \!\!\! \di^4 x' \;  
  \langle [\widehat O(x),\wj^\mu(x')] \rangle_\subb \e^{-\ii K \cdot (x'-x)}  
\eea
where $K_T$ is the projection of $K$ orthogonal to $n$. This, as it will become clear
later, is the covariant form of the same formula obtained in ref.~\cite{hosoya}, 
with the (important) addition of the current term. In other words, it is the well 
known formula expressing the transport coefficients as frequence derivatives of 
the retarded correlators of stress-energy components, the so-called Kubo formula. 
Defining:
\bea\label{retg}
  (\widehat X, \widehat Y) \equiv &&
  n^\alpha \dfrac{\partial}{\partial K^\alpha}\Big|_{n \cdot k=0} \lim_{k_T \to 0} 
  {\rm Im} \; \ii T \int_{-\infty}^{t} \!\!\! \di^4 x^\prime \;  
  \langle [\widehat X(x),\widehat Y(x')] \rangle_{\subb} \e^{-\ii K \cdot (x'-x)}  
  \nonumber \\
 = && n^\alpha \dfrac{\partial}{\partial K^\alpha}\Big|_{n \cdot k=0} \lim_{k_T \to 0} 
  {\rm Im} \; \ii T \int_{-\infty}^{0} \!\!\! \di^4 x^\prime \;  
  \langle [\widehat X(0),\widehat Y(x')] \rangle_{\subb} \e^{-\ii K \cdot x'}
\eea
which is bilinear in $\widehat X$ and $\widehat Y$, one can write the deviations 
of the stress-energy tensor from its LTE value as:
\be\label{set1}
  \langle \wT^{\mu\nu}(x) \rangle - \langle \wT^{\mu\nu}(x) \rangle_{\rm LE}
  \equiv \delta T^{\mu\nu}(x) \simeq (\wT^{\mu\nu},\wT^{\rho\sigma}) \, 
  \partial_\rho \beta_\sigma(x) - (\wT^{\mu\nu},\wj^\rho)\, \partial_\rho \zeta(x)
\ee  
Similarly, the deviation of the current with respect to its value at LTE reads:
\be\label{curr1}
  \langle \wj^{\mu}(x) \rangle - \langle \wj^{\mu}(x) \rangle_{\rm LE}
  \equiv \delta j^{\mu}(x) = (\wj^{\mu},\wT^{\rho\sigma}) \, 
  \partial_{\rho}\beta_{\sigma}(x) - (\wj^{\mu},\widehat{j}^{\rho}) \, 
  \partial_\rho\zeta(x)
\ee

The next step is to decompose the correlators and the gradients of the relativistic
fields into irreducible components under rotations, a procedure leading to the 
identification of the familiar transport coefficients: shear and bulk viscosities,
thermal conductivities etc. We are not going to show how this is accomplished, we
would just like to point out, for the purpose of the identification of the transport
coefficients, that the gradients of $\beta$ can be turned into the gradients of 
the velocity field $u$ by using (\ref{ubeta}). Having defined:
$$
  \Delta_{\mu\nu} = g_{\mu\nu} - u_\mu u_\nu
$$
and
$$
 D= u \cdot \partial \qquad \qquad \nabla_T^\mu = \partial^\mu - u^\mu D
$$
the transverse gradients of the velocity field $\nabla_T^\mu u^\nu$ can 
be written as follows:
\bea
  \nabla_{T\mu} u^\nu && = \nabla_{T\mu} \frac{\beta^\nu}{\sqrt{\beta^2}} = \beta^\nu
 \left( -\frac{1}{2}\right) (\beta^2)^{-3/2} \nabla_{T\mu} \beta^2 + \frac{1}{\sqrt{\beta^2}}
 \nabla_{T\mu} \beta^\nu \nonumber \\
 && = \frac{1}{\sqrt{\beta^2}} \left( - \frac{\beta^\nu \beta^\rho}{\beta^2} 
 \nabla_{T\mu} \beta_\rho + \nabla_{T\mu} \beta^\nu \right) = 
 \frac{1}{\sqrt{\beta^2}} \Delta^{\rho\nu} \nabla_{T\mu} \beta_\rho,
\eea
where we have used the relation (\ref{ubeta}). Thereby, the Navier-Stokes shear term
can be fully expressed in terms of the inverse temperature four-vector $\beta$ and
its gradients. The same transformation can be proved for the other terms \cite{betaframe}.

\section{Outlook}   

The non-equilibrium statistical operator method introduced by D. Zubarev more than 
50 years ago has been a very important achievement in statistical physics, which 
has not received the deserved attention. It can be used in all physical problems 
where local thermodynamic equilibrium is reached and it can be quite straightforwardly
extended to relativistic statistical mechanics. In this work, we have just presented 
an amendment of its original formulation which reproduces known results and makes 
its application easier to relativistic hydrodynamics problems. 
Since it is a fully-fledged quantum framework, this approach is especially suitable 
for the calculation of quantum effects. Amongst various applications, the recent 
evidence of non-vanishing polarization in the Quark Gluon Plasma \cite{starnat}, 
makes it the ideal tool to deal with this newly found phenomenon.



\section*{APPENDIX A - Supplementary notes on the derivation of the Kubo formula}

Working out the eq.~(\ref{kuboexp}) requires the eqs.~(\ref{aa}) and (\ref{bb}).
By also using the approximations (\ref{ltetoeq}) and (\ref{cartesianize}), the 
(\ref{kuboexp}) turns into: 
\bea\label{bint}
  \langle \widehat O (x) \rangle - \langle \widehat O (x) \rangle_{\rm LE} & \simeq &
  -\int_{t_0}^{t} \di t^\prime \; \int \di^3 \x' \; \langle \widehat O (x) 
  \rangle_{\rm LE} \left( \langle \wT^{\mu\nu}(t',{\bf x}') \rangle_{\rm LE} 
  \partial_\mu \beta_\nu - \langle \wj^\mu (t',{\bf x}') \rangle_{\rm LE} 
  \partial_\mu \zeta \right) \nonumber \\
  &+& \int_0^1 \di z \; \int_{t_0}^{t} \di t^\prime \; \int \di^3 \x' \;
  \left( \langle \widehat O (x) \e^{ -z (\beta(x) \cdot \widehat P  \zeta(x) \widehat Q)} 
  \wT^{\mu\nu}(t',{\bf x}') \e^{ z (\beta(x) \cdot \widehat P - \zeta(x) \widehat Q)} 
  \rangle_{\rm LE} \partial_\mu \beta_\nu \right. \nonumber \\
  & - & \left. \langle \widehat O (x) \e^{ -z (\beta(x) \cdot \widehat P - \zeta(x) \widehat Q)} 
  \wj^{\mu}(t',{\bf x}') \widehat \e^{ z (\beta(x) \cdot \widehat P - \zeta(x) \widehat Q)} 
  \rangle_{\rm LE} \partial_\mu \zeta \right)
\eea
Being $[\widehat Q,\wT(x)]=0$ and $[\widehat Q,\wj(x)]=0$ one can also write:
$$
    \e^{ -z (\beta(x) \cdot P - \zeta(x) \widehat Q)} \widehat X(t',{\bf x}') 
    \, \e^{ z (\beta(x) \cdot P - \zeta(x) \widehat Q)} = 
    \widehat X(t'+iz\sqrt{\beta^2},{\bf x}')
$$
with $\widehat X = \wT, \wj$, where, in the last expression we have tacitly assumed
that $n = \hat \beta$, i.e. that the local equilibrium hypersurface coincides - 
locally around $x$ - with the hypersurface normal to $\beta$ \cite{betaframe}. 
Hence, the last term in the right hand side of eq.~(\ref{bint}) can be rewritten as: 
$$
  \int_0^1 \di z \; \int_{t_0}^{t} \di t' \; \int \di^3 \x' \;
  \left( \langle \widehat O (x) \wT^{\mu\nu}(t'+\ii z\sqrt{\beta^2},{\bf x}') 
  \rangle_{\rm LE} \partial_\mu \beta_\nu - \langle \widehat O (x) 
  \wj^{\mu}(t' + \ii z \sqrt{\beta^2},{\bf x}') \rangle_{\rm LE} \partial_\mu 
  \zeta \right)
$$
provided that $n = \hat \beta$, that is if the local equilibrium hypersurface 
coincides - locally - with the hypersurface normal to $\beta$ \cite{betaframe}.
The operator $\widehat X = \wT, \wj$, can be rewritten:
\bea\label{} 
  \widehat X (t'+\ii z\sqrt{\beta^2},{\bf x}') &=& \widehat X (t_0+\ii z 
  \sqrt{\beta^2},{\bf x}') + \int_{t_0}^{t'} \di \theta \; \frac{\partial}{\partial \theta}
  \widehat X (\theta+\ii z \sqrt{\beta^2},{\bf x}') \nonumber \\
  &=& \widehat X (t_0+\ii z \sqrt{\beta^2},{\bf x}') + \int_{t_0}^{t'} \di 
  \theta \; \frac{1}{\ii\sqrt{\beta^2}} 
  \frac{\partial}{\partial z} \widehat X (\theta+\ii z \sqrt{\beta^2},{\bf x}')
\eea  
Integrating in $z$:
\bea\label{zint}
 && \int_0^1 \di z \langle \widehat O (x) \widehat X(t'+\ii z\sqrt{\beta^2},{\bf x}') 
  \rangle_{\rm LE} = \int_0^1 \di z \; \langle \widehat O(x) \widehat X 
  (t_0+\ii z \sqrt{\beta^2},{\bf x}') \rangle_{\rm LE} \nonumber \\
 && + \int_{t_0}^{t'} \di \theta \; 
  \frac{1}{\ii\sqrt{\beta^2}} \left( \langle \widehat O (x) \widehat X 
  (\theta+ \ii \sqrt{\beta^2},{\bf x}') \rangle_{\rm LE} -
  \langle \widehat O (x) \widehat X (\theta,{\bf x}') \rangle_{\rm LE} \right) 
\eea
Now we use the same approximation of $\widehat A$ like in (\ref{ltetoeq}) and 
the LTE mean values are calculated at equilibrium, with the density operator 
(\ref{eqlbrm}). Thus:
\begin{eqnarray*} 
 &&\langle \widehat O (x) \widehat X (\theta+ \ii \sqrt{\beta^2},{\bf x}') \rangle_{\rm LE} -
 \langle \widehat O (x) \widehat X (\theta,{\bf x}') \rangle_{\rm LE} 
 \simeq \langle \widehat O (x) \widehat X (\theta+ \ii \sqrt{\beta^2},{\bf x}') 
 \rangle_\subb - \langle \widehat O (x) \widehat X (\theta,{\bf x}') \rangle_\subb 
 \nonumber \\
 && = \langle \widehat O (x) \e^{ -\beta(x) \cdot \widehat P + \zeta(x) \widehat Q)} 
 \widehat X (\theta,{\bf x}') \e^{ \beta(x) \cdot \widehat P - \zeta(x) \widehat Q)} 
 \rangle_\subb - \langle \widehat O (x) \widehat X (\theta,{\bf x}') \rangle_\subb
 \nonumber \\
 && = \langle \widehat X (\theta,{\bf x}') \widehat O (x) \rangle_\subb - 
 \langle \widehat O (x) \widehat X (\theta,{\bf x}') \rangle_\subb = 
 \langle [\widehat X (\theta,{\bf x}'), \widehat O (x)] \rangle_\subb 
\end{eqnarray*}
Substitution into the (\ref{zint}) yields:
\be\label{zint2}
 \int_0^1 \di z \langle \widehat O (x) \widehat X(t+\ii z\sqrt{\beta^2},{\bf x}') 
  \rangle_{\rm LE} \simeq \int_0^1 \di z \; \langle \widehat O(x) \widehat X 
  (t_0+\ii z \sqrt{\beta^2},{\bf x}') \rangle_\subb + 
  \frac{1}{\ii\sqrt{\beta^2}} \int_{t_0}^{t'} \di 
  \theta \; \langle [\widehat X (\theta,{\bf x}'), \widehat O (x)] \rangle_\subb 
\ee
Using this result for $\widehat X = \wT,\wj$ allows to turn the (\ref{bint}) into:
\bea\label{bint2}
  \langle \widehat O (x) \rangle - \langle \widehat O (x) \rangle_{\rm LE} & \simeq &
  \int_{t_0}^{t} \di t' \; \int \di^3 \x' \; \int_0^1 \di z \; \left[
  \left( \langle \widehat O(x) \wT^{\mu\nu} (t_0+\ii z \sqrt{\beta^2},{\bf x}') \rangle_\subb
  -  \langle \widehat O (x) \rangle_\subb \langle \wT^{\mu\nu}(t',{\bf x}') \rangle_\subb
  \right) \partial_\mu \beta_\nu \right. \nonumber \\ 
  && \left. \left( \langle \widehat O(x) \wj^\mu (t_0+\ii z \sqrt{\beta^2},{\bf x}') \rangle_\subb
  -  \langle \widehat O (x) \rangle_\subb \langle \wj^{\mu}(t',{\bf x}') \rangle_\subb
  \right)\partial_\mu \zeta \right]
  \nonumber \\
  &+& \ii T \int_{t_0}^{t} \di t' \; \int_{t_0}^{t'} \di \theta 
  \int \di^3 \x' \; \left(\langle [\widehat O(x),\wT^{\mu\nu} (\theta,{\bf x}')] \rangle_\subb 
  \partial_\mu \beta_\nu(x') - \langle [\widehat O(x),\wj^\mu(\theta,{\bf x}')] \rangle_\subb
  \partial_\mu \zeta(x') \right)  
\eea

In the paper by A.~Hosoya {\it et al.} \cite{hosoya}, in the limit $t_0 \to -\infty$, 
the first of the two integral terms is shown to be vanishing, based on the idea that
$\lim_{t_0 \to -\infty} \widehat X (t_0+\ii z \sqrt{\beta^2},{\bf x}') \simeq 
\lim_{t_0 \to -\infty} \widehat X (t_0,{\bf x}')$ and that correlation between an
operator $\widehat O$ at time $t$ and $\widehat X$ at an infinitely remote past is
0, that is:
$$
 \lim_{t_0 \to -\infty} \langle \widehat O(x) \wT^{\mu\nu} (t_0,{\bf x}') \rangle_\subb
 \simeq \lim_{t_0 \to -\infty} \langle \widehat O(x) \rangle_\subb \langle \wT^{\mu\nu} 
 (t_0,{\bf x}') \rangle_\subb = \langle \widehat O(x) \rangle_\subb \langle \wT^{\mu\nu} 
 (t',{\bf x}') \rangle_\subb
$$
where in the last equality we have taken advantage of the fact that the mean value 
of any operator is constant at equilibrium. Therefore, the first integral on the
right hand side of eq.~(\ref{bint2}) vanishes and we are only left with the 
second integration, that is eq.~(\ref{ltexpa}). 
 
\end{document}